\documentclass[lettersize,journal]{IEEEtran}
\usepackage{amsmath,amssymb,amsfonts,amsthm,bm}
\usepackage{graphicx}
\usepackage[caption=false,font=footnotesize]{subfig}
\usepackage{booktabs}
\usepackage{multirow}
\usepackage{cite}
\usepackage{algorithm}
\usepackage{algorithmic}
\usepackage{xcolor}
\usepackage{balance}
\usepackage[colorlinks=true,linkcolor=red,citecolor=blue,urlcolor=blue]{hyperref}

\newcommand{\E}{\mathbb{E}}
\newcommand{\C}{\mathbb{C}}
\newcommand{\R}{\mathbb{R}}

\newcommand{\mB}{\mathcal{B}}
\newcommand{\mS}{\mathcal{S}}
\newcommand{\mL}{\mathcal{L}}

\graphicspath{{./}}
\begin{document}
\title{Learned Blockwise Port Activation for Real Time Beamforming in Fluid Antenna Arrays}

\author{Yuanhui Wu, Zhentian Zhang, 
	Hanjiang Hong, 
	Hao Jiang,~\IEEEmembership{Senior Member,~IEEE},
	\\Zaichen Zhang,~\IEEEmembership{Senior Member,~IEEE}, 
	Kai-Kit Wong,~\IEEEmembership{Fellow,~IEEE},
	Yin Xu~\IEEEmembership{Senior Member,~IEEE},
	\\and Wenjun Zhang,~\IEEEmembership{Fellow,~IEEE}, 
	\thanks{Yuanhui Wu is with the College of Artificial Intelligence, Nanjing University of Information Science and Technology, 210044, P. R. China (e-mails: 202412621447@nuist.edu.cn).}%
	\thanks{Zhentian Zhang is with The Hong Kong Polytechnic University, Hong Kong SAR, China. (e-mail: zhentianzhangzzt@gmail.com).}%
	\thanks{H. Jiang is with the School of Cyber Science and Engineering, Southeast University, Nanjing 210096, P. R. China (e-mail: jiang.hao@seu.edu.cn).}
	\thanks{Zaichen Zhang is with the National Mobile Communications Research Laboratory, Frontiers Science Center for Mobile Information Communication and Security, Southeast University, Nanjing, 210096, China and also with the Purple Mountain Laboratories, Nanjing 211111, China (e-mail: zczhang@seu.edu.cn).}
	\thanks{Kai-Kit Wong and Hanjiang Hong are with the Department of Electronic and Electrical Engineering, University College London, Torrington Place, WC1E 7JE, United Kingdom  (e-mail: \{kai-kit.wong, hanjiang.hong\}@ucl.ac.uk). }
	\thanks{Yin Xu and Wenjun Zhang are with the Cooperative Medianet Innovation Center (CMIC), Shanghai Jiao Tong University, Shanghai, China (e-mail: \{xuyin, zhangwenjun\}@sjtu.edu.cn).}
	\thanks{Corresponding Author: Hao Jiang}
}

%

\maketitle

\begin{abstract}
	Fluid antenna arrays (FAAs), support multiuser downlink transmission by activating a subset of reconfigurable ports. The activation mask jointly determines the effective channel and the sparse radiating aperture, which requires a balance among sum rate, sidelobe suppression, hardware constraints, and online complexity. Channel driven selection can cluster active ports and increase sidelobes, whereas sidelobe oriented synthesis is typically channel independent and can sacrifice sum rate. This paper proposes learned blockwise port activation (L-BPA), for real time sidelobe aware FAA downlink beamforming. L-BPA activates a fixed number of ports in each aperture block, which supports grouped switching hardware and limits port clustering. A lightweight convolutional network scores ports using multiuser channel features, port coordinates, and user power statistics. Training combines blockwise straight through masks with a differentiable peak sidelobe level (PSLL), surrogate. During inference, learned scores are combined with multiscale geometric repulsion, followed by regularized zero forcing precoding over the reduced effective channel. L-BPA reduces the average PSLL by 3.26 dB relative to uniform sparse activation while achieving a slightly higher sum rate. It also reduces the PSLL by 8.13 dB and 10.10 dB relative to greedy and gain based selection, respectively, without iterative online search.
\end{abstract}


\IEEEpeerreviewmaketitle

\section{Introduction}

Fluid antenna systems (FASs) have emerged as flexible antenna architectures in which active radiating elements can be repositioned or selected within a finite physical aperture \cite{wong2020_fas_limits,Wong2021FAS}. This antenna level spatial reconfigurability provides a design dimension beyond conventional architectures with fixed antenna locations. Early FAS studies primarily exploited spatial channel variations within the available aperture, commonly termed fading diversity \cite{r1,r2,r3,wt0,wt1,wt2,wt3,r4,r5.1,r5.2,r5.3,r5.4,r5.5,r5.6,r6}. By selecting favorable antenna positions, fading diversity can improve signal strength and transmission reliability while mitigating interference.

Fluid antenna arrays (FAAs) extend this concept by jointly configuring multiple active radiating positions \cite{FAA1,FAA2}. Beyond channel fluctuations, the selected port positions determine the effective aperture, spatial sampling pattern, interelement spacings, and array manifold. These properties directly govern the radiation pattern and spatial frequency response. This controllable spatial capability is termed \emph{geometry diversity}. Unlike fading diversity, which arises from the random propagation environment, geometry diversity is determined by the configurable spatial structure of the array. It can therefore be predicted, optimized, and adapted to the communication and radiation objectives.

Appropriate port activation enables an FAA to reshape its radiation characteristics and improve beamforming gain, directivity control, and sidelobe suppression \cite{beam1,beam2,beam3}. It can also enable the exploitation of mutual coupling effects \cite{beam3,Ramirez2025MetasurfaceFAS,Ramirez2026BeamspaceFAS}. These capabilities can reduce the dependence on additional radio frequency chains, phase shifters, and delay elements used in conventional beamforming architectures. However, practical FAAs operate within a finite aperture, where performance depends on the interaction among aperture size, port placement, spatial sampling, and sidelobe behavior \cite{FAA1,FAA2}. {\em FAA port activation is therefore a geometry-aware array design problem rather than only a channel selection problem}.

In multiuser downlink transmission, the binary activation mask determines the effective channel matrix used by the precoder. The same mask also defines the sparse physical aperture that radiates the downlink signals. Each activation decision consequently affects two coupled quantities: the regularized zero forcing (RZF), sum rate and the sidelobe characteristics of the sparse aperture. Channel gain and rate driven selectors can exploit instantaneous channel state information and identify ports that are favorable for multiuser transmission. However, they can concentrate active ports in spatially favorable regions. This clustering reduces the effective aperture and can increase the peak sidelobe level (PSLL). In contrast, sparse array synthesis methods can generate well distributed port configurations with lower PSLL. These methods are usually channel independent and can exclude ports that are important for channel gain and multiuser spatial separation.

The tradeoff becomes more challenging in Internet of Things oriented FAA downlink systems. The transmitter must generate a binary activation mask at the channel time scale, satisfy a strict active port budget, and remain compatible with grouped switching hardware. Exhaustive subset search is infeasible when candidate ports form a dense planar grid. Iteratively evaluating RZF precoders over many candidate masks also introduces excessive online complexity.

Antenna selection has been widely studied to reduce the number of radio frequency chains in conventional multiple input multiple output systems \cite{Sanayei2004AntennaSelection}. FAS port selection has also been investigated for identifying favorable fluid antenna positions from limited channel observations \cite{Chai2022PortSelection}. More recent studies have considered joint port selection and beamforming \cite{Zhang2024JointPortBF,Chen2026NearField}, finite aperture FAA geometry design \cite{FAA1,FAA2}, PSLL suppression for planar sparse FAAs \cite{beam2}, and channel acquisition over dense FAS ports \cite{Zhang2026JADCE,Zhang2026GSCR}. However, these studies do not directly provide a {\em real time} and {\em hardware compatible} activation strategy that is simultaneously channel adaptive and sidelobe aware for multiuser FAA downlink precoding.

This paper addresses this gap by proposing learned blockwise port activation (L-BPA). The proposed design follows a hybrid learning and deterministic selection framework. Instead of directly generating an unconstrained binary mask, a lightweight neural network assigns importance scores to candidate ports using multiuser channel features. A deterministic blockwise selector then enforces the active port budget and grouped switching constraints. During training, blockwise straight through masks align the optimization objective with the discrete inference constraint. A differentiable PSLL surrogate discourages activation patterns with high sidelobes. During inference, a multiscale geometry aware rule combines learned channel scores with local repulsion among nearby ports. The resulting feasible port set is directly applied to a reduced dimensional RZF precoder without iterative online search.

\subsection{Contributions}

The main contributions are summarized as follows.

\begin{itemize}
	\item The multiuser FAA downlink design is formulated as a \emph{block-wise} sparse port activation problem subject to a strict active-port budget, where the block-wise structure significantly alleviates the computational complexity of the underlying NP-hard activation problem. Each activation mask is evaluated using both RZF sum rate and broadside PSLL, which explicitly captures the tradeoff between communication performance and sidelobe suppression.
	
	\item L-BPA combines lightweight channel aware port scoring with constrained discrete selection. Blockwise straight through masks and a differentiable PSLL surrogate are used during training. A multiscale geometry aware selector is used during inference to limit port clustering without iterative RZF search.
	
	\item Extensive simulations compare L-BPA with full port, structured sparse, channel driven, greedy, and PSLL oriented baselines. The evaluation includes sum rate distributions, beam patterns, port layouts, activation tradeoffs, generalization without retraining, ablation studies, and online complexity.
\end{itemize}

The remainder of this paper is organized as follows. Section~\ref{sec.2} presents the system model and performance metrics. Section~\ref{sec.3} describes L-BPA and analyzes its online complexity. Section~\ref{sec.4} presents the simulation setup and numerical results. Section~\ref{sec.5} concludes the paper. {\em notation:} Scalars, vectors, matrices, and sets are denoted by italic, boldface lowercase, boldface uppercase, and calligraphic letters, respectively. The superscripts $(\cdot)^T$ and $(\cdot)^H$ denote transpose and Hermitian transpose. Moreover, $\|\cdot\|_0$, $\|\cdot\|$, and $\|\cdot\|_F$ denote the $\ell_0$, Euclidean, and Frobenius norms, respectively. $\mathbb E[\cdot]$ denotes expectation and $\mathbb R$ and $\mathbb C$ denote the real and complex fields.
\begin{figure*}[t]
	\centering
	\includegraphics[width=\textwidth]{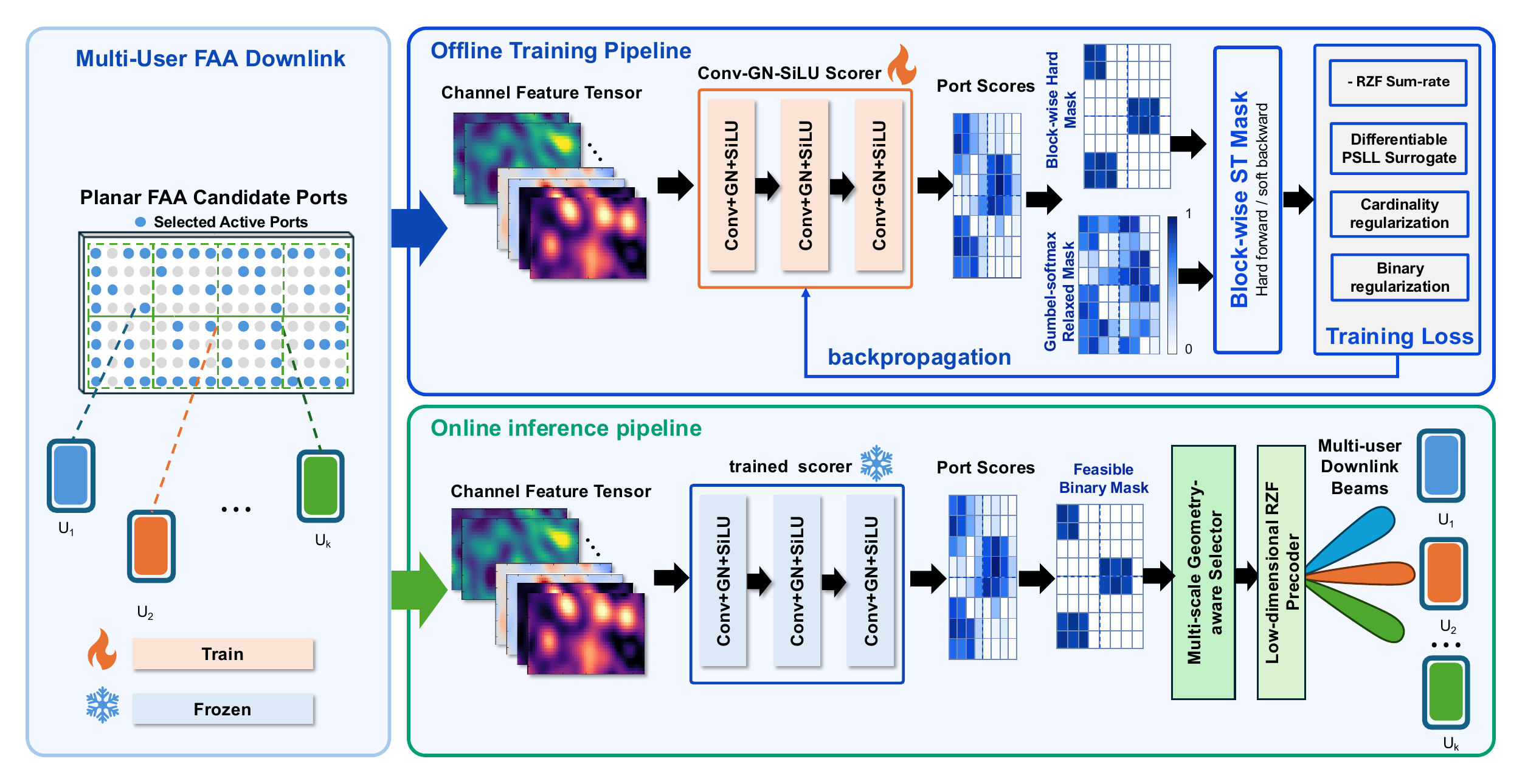}
	\caption{Overall workflow of the proposed L-BPA method. The left panel corresponds to the multi-user FAA downlink model in Section~\ref{sec.2}, where a dense planar candidate-port set is used to serve multiple users with a limited active-port budget. The upper branch shows the offline training procedure: the full-aperture CSI is converted into the channel feature tensor and processed by the Conv--GN--SiLU scorer in Section~\ref{subsec:feature_scorer}, and the resulting logits are converted into block-wise straight-through masks in Section~\ref{subsec:st_training}. The training objective combines the RZF sum-rate, differentiable PSLL surrogate, cardinality regularization, and binary regularization. The lower branch shows online inference: the trained scorer is frozen, port scores are generated for a new CSI realization, the multi-scale geometry-aware selector in Section~\ref{subsec:geometry_inference} produces a feasible binary mask, and low-dimensional RZF precoding is then applied on the selected ports.}
	\label{fig:lbpa_framework}
\end{figure*}
\section{System Model and Problem Formulation}\label{sec.2}

\subsection{FAA Downlink Model}

Consider an FAA access point serving $K$ single-antenna users in the downlink. The FAA contains $N=N_xN_y$ candidate ports over a rectangular aperture. The normalized position of port $n$ is $\mathbf p_n=[x_n,y_n]^T$, where coordinates are measured in wavelengths. At one scheduling interval, only $M$ ports are activated. The activation vector is
\begin{equation}
	\mathbf s=[s_1,\ldots,s_N]^T\in\{0,1\}^{N},~\|\mathbf s\|_0=M .
\end{equation}
Let $\mS=\{n:s_n=1\}$ be the selected port set.

The full-aperture channel of user $k$ is $\mathbf h_k\in\C^N$. Each path is mapped to a direction-cosine pair $(u_{k,\ell},v_{k,\ell})$ and quantized onto a finite visible angular grid. Let $(u^{\rm q}_{k,\ell},v^{\rm q}_{k,\ell})=\mathcal Q(u_{k,\ell},v_{k,\ell})$ denote the nearest point on a $N_u\times N_v$ grid over $[-1,1]^2$ satisfying $(u^{\rm q}_{k,\ell})^2+(v^{\rm q}_{k,\ell})^2\le 1$. The finite-port channel is generated as
\begin{equation}
	h_{k,n}=\sum_{\ell=1}^{L_k}\alpha_{k,\ell}
	\exp\left(j2\pi(x_nu^{\rm q}_{k,\ell}+y_nv^{\rm q}_{k,\ell})\right),
	\label{eq:channel}
\end{equation}
where $\alpha_{k,\ell}$ is the complex path gain. The discrete angular response is used only to generate finite-port CSI. L-BPA operates on the resulting multi-user CSI tensor and does not require path parameters. After channel generation, each user's port-domain channel is root mean square (RMS)-normalized over the aperture, so the reported behavior reflects spatial port selection rather than large-scale path-loss imbalance. The selected effective channel is
\begin{equation}
	\mathbf H_{\mS}=\left[\mathbf h_1[\mS]^T;\ldots;\mathbf h_K[\mS]^T\right]\in\C^{K\times M}.
\end{equation}

\subsection{RZF Beamforming Objective}

For a given selected set $\mS$, the base station computes the regularized zero-forcing (RZF) precoder \cite{Wagner2012RZF}
\begin{equation}
	\mathbf P_{\mS}=\mathbf H_{\mS}^{H}\left(\mathbf H_{\mS}\mathbf H_{\mS}^{H}+\eta_{\rm rzf}\mathbf I_K\right)^{-1},
	\label{eq:rzf}
\end{equation}
followed by total-power normalization. Let $\widetilde{\mathbf P}_{\mS}$ be the normalized precoder satisfying $\|\widetilde{\mathbf P}_{\mS}\|_F^2=1$ and
\begin{equation}
	\mathbf G_{\mS}=\mathbf H_{\mS}\widetilde{\mathbf P}_{\mS}
\end{equation}
be the effective user coupling matrix. At desired SNR $\rho$, the SINR of user $k$ is
\begin{equation}
	\gamma_k(\mathbf s)=
	\frac{\rho |[\mathbf G_{\mS}]_{k,k}|^2}
	{1+\rho\sum_{j\ne k}|[\mathbf G_{\mS}]_{k,j}|^2},
	\label{eq:sinr}
\end{equation}
and the downlink sum-rate is
\begin{equation}
	R(\mathbf s)=\sum_{k=1}^{K}\log_2(1+\gamma_k(\mathbf s)).
	\label{eq:rate}
\end{equation}
The communication-driven activation problem is
\begin{align}
	\max_{\mathbf s}\quad & R(\mathbf s)\\
	\text{s.t.}\quad & \mathbf s\in\{0,1\}^{N},\quad \|\mathbf s\|_0=M .
	\label{eq:comb_problem}
\end{align}
The feasible set has combinatorial size and direct subset search is infeasible for dense FAA grids.

\subsection{Broadside PSLL Metric}

Sparse activation also changes the physical aperture distribution. To quantify sidelobe behavior, we evaluate the equal-amplitude broadside array factor
\begin{equation}
	AF_{\mathbf s}(u,v)=\sum_{n=1}^{N}s_n
	\exp\left(j2\pi(x_nu+y_nv)\right).
	\label{eq:af}
\end{equation}
The normalized beam power is $|AF_{\mathbf s}(u,v)|^2/M^2$. Let $A_x$ and $A_y$ denote the aperture side lengths in wavelengths. The sidelobe region used in the experiments is
\begin{equation}
	\mathcal G_{\rm sl}=\{(u,v):u^2+v^2\le1,\ |u|\ge A_x^{-1}\ {\rm or}\ |v|\ge A_y^{-1}\}.
	\label{eq:psll_guard}
\end{equation}
This excludes the broadside guard rectangle determined by the aperture dimensions. The PSLL is the maximum normalized beam power over $\mathcal G_{\rm sl}$. This equal-amplitude PSLL is not the RZF communication objective; it is a geometry proxy for sparse-aperture sidelobes. The actual user beams after RZF depend on data-stream weights and user channels, whereas this metric isolates the sidelobe tendency introduced by the selected port layout itself.

The rate--PSLL conflict can be written explicitly by defining
\begin{equation}
	P_{\rm sl}(\mathbf s)=
	\max_{(u,v)\in\mathcal G_{\rm sl}}
	\frac{|AF_{\mathbf s}(u,v)|}{\sum_n s_n}.
	\label{eq:psll_amp}
\end{equation}
A channel-only selector tends to solve \eqref{eq:comb_problem}, which may place many ports in a high-gain local region and increase $P_{\rm sl}(\mathbf s)$. A pattern-only sparse synthesis instead solves a channel-independent problem such as $\min_{\mathbf s}P_{\rm sl}(\mathbf s)$, which can waste ports from the viewpoint of the instantaneous multi-user channel. The desired sparse FAA operating point is therefore the constrained bi-objective problem
\begin{align}
	\max_{\mathbf s}\quad & R(\mathbf s)\\
	\text{s.t.}\quad & \mathbf s\in\{0,1\}^{N},\quad \|\mathbf s\|_0=M,\quad
	P_{\rm sl}(\mathbf s)\le P_0 ,
	\label{eq:rate_psll_tradeoff}
\end{align}
or, equivalently, a scalarized form $R(\mathbf s)-\lambda_{\rm sl}[P_{\rm sl}(\mathbf s)-P_0]_+$. L-BPA follows this scalarized view: the neural scorer learns the channel-dependent rate term, the block constraint restricts the feasible hardware set, the differentiable PSLL surrogate provides gradients for the sidelobe constraint during training, and the inference-stage repulsion discourages local port clustering before the final binary mask is passed to RZF.

\section{Proposed learned block-wise port activation}\label{sec.3}

Fig.~\ref{fig:lbpa_framework} summarizes the correspondence between the system model, offline training, and online inference. L-BPA consists of four modules: the block-wise hardware constraint in Section~\ref{subsec:block_constraint}, the channel feature tensor and Conv--GN--SiLU scorer in Section~\ref{subsec:feature_scorer}, the block-wise straight-through training mask in Section~\ref{subsec:st_training}, and the multi-scale geometry-aware inference rule in Section~\ref{subsec:geometry_inference}.

\subsection{Block-Wise Hardware Constraint}\label{subsec:block_constraint}

The FAA aperture is divided into $B_x\times B_y$ blocks. Let $\mB_b$ be the port index set in block $b$, and let $m_b$ be the number of ports activated in each block. The feasible set becomes
\begin{equation}
	\sum_{n\in\mB_b}s_n=m_b,~ b=1,\ldots,B_xB_y,
	\label{eq:block_constraint}
\end{equation}
with $M=B_xB_y m_b$. This constraint matches block-level switch control and prevents all selected ports from collapsing into a small high-gain area.

\subsection{Channel Feature Tensor and Neural Scorer}\label{subsec:feature_scorer}

For each realization, the input to the neural scorer is a full-aperture tensor. Let $\bar h_n=K^{-1}\sum_k h_{k,n}$, $d_n=(K^{-1}\sum_k|h_{k,n}-\bar h_n|^2)^{1/2}$, $p_n^{\rm u}=K^{-1}\sum_k |h_{k,n}|^2$, and $q_n=|\bar h_n|^2/(1+d_n^2)$. The nonconstant port statistics used by the implemented feature tensor are summarized as
\begin{equation}
	\tilde{\mathbf x}_n=\mathcal N\!\left([\mathbf c_n^T,\mathbf r_n^T,\mathbf u_n^T]^T\right),
	\label{eq:feature}
\end{equation}
where
\begin{equation}
	\begin{aligned}
		\mathbf c_n&=[x_n,y_n]^T,\\
		\mathbf r_n&=[\Re\bar h_n,\Im\bar h_n,|\bar h_n|,
		d_n,|\bar h_n|^2,d_n^2,\\
		&\quad q_n,\log(q_n+\epsilon),\angle\bar h_n]^T,\\
		\mathbf u_n&=[p_n^{\rm u},\operatorname{std}_k(|h_{k,n}|^2)]^T .
	\end{aligned}
\end{equation}
The operator $\mathcal N(\cdot)$ denotes per-realization feature standardization over the aperture. In the implementation, the real dispersion proxy $d_n$ is stored in the same real--imaginary--magnitude template as complex channel features; after this template expansion and the two user-power statistics in $\mathbf u_n$, the actual input tensor has 15 channels. This bookkeeping does not introduce an additional selection rule. Stacking the expanded port features over the $N_x\times N_y$ grid gives $\mathbf X\in\R^{15\times N_x\times N_y}$.

A lightweight convolutional scorer maps $\mathbf X$ to port logits. With $\phi(\cdot)$ denoting SiLU and ${\rm GN}(\cdot)$ denoting group normalization \cite{Wu2018GroupNorm}, its backbone is
\begin{equation}
	\mathbf F^{(l)}=\phi\!\left({\rm GN}\left(\mathbf W^{(l)}*\mathbf F^{(l-1)}\right)\right),~ l=1,2,3,
	\label{eq:cnn}
\end{equation}
where $\mathbf F^{(0)}=\mathbf X$ and $*$ denotes a $3\times3$ convolution. A $1\times1$ convolution gives raw logits $\tilde{\mathbf z}$. The final scoring rule is
\begin{equation}
	\mathbf z=f_{\bm\theta}(\mathbf X)
	=a_\theta\operatorname{vec}\!\left(\mathbf W_{\rm h}*\mathbf F^{(3)}\right)
	+b_\theta\mathbf q\in\R^N ,
	\label{eq:scorer}
\end{equation}
where $\mathbf q=[q_1,\ldots,q_N]^T$ is the normalized analytic-score channel and $a_\theta,b_\theta$ are learned scalar adapter parameters. The network therefore acts only as a channel-aware scorer; the final binary mask is generated by an explicit feasible selector.

\subsection{Block-Wise Straight-Through Training}\label{subsec:st_training}

Training uses the same block-wise selection structure as inference. For block $b$, a soft top-$m_b$ mask is formed from logits $\mathbf z_b=\{z_n:n\in\mB_b\}$ using a Gumbel-softmax relaxation \cite{Jang2017GumbelSoftmax},
\begin{equation}
	\mathbf a_b^{\rm soft}=\operatorname{clip}\left(m_b\operatorname{softmax}\left(\frac{\mathbf z_b+\mathbf g_b}{\tau}\right),0,1\right),
\end{equation}
where $\mathbf g_b$ is Gumbel noise and $\tau$ is the temperature. The hard block mask is $\mathbf a_b^{\rm hard}=\operatorname{TopM}(\mathbf z_b,m_b)$. Following the straight-through gradient estimator \cite{Bengio2013STE}, the block mask is
\begin{equation}
	\mathbf a_b=\mathbf a_b^{\rm hard}+\mathbf a_b^{\rm soft}-\operatorname{stopgrad}(\mathbf a_b^{\rm soft}).
	\label{eq:st_mask}
\end{equation}
Concatenating all block masks gives a forward mask that exactly satisfies \eqref{eq:block_constraint}.

The selected channel used in the differentiable RZF backend is $\mathbf H_{\mathbf a}=\mathbf H\operatorname{diag}(\mathbf a)/\sqrt{\sum_n a_n+\epsilon}$. To make sidelobe suppression part of learning rather than only a post-processing criterion, we also use a differentiable broadside PSLL surrogate. Let $\mathcal{G}_{\rm sl}$ denote the visible $u$--$v$ samples outside the broadside mainlobe guard region. For a training mask $\mathbf a$, define
\begin{equation}
	\widehat P_{\rm sl}(\mathbf a)=
	\frac{1}{\beta}\log\sum_{(u,v)\in\mathcal{G}_{\rm sl}}
	\exp\left(
	\beta\frac{|AF_{\mathbf a}(u,v)|}{\sum_n a_n+\epsilon}
	\right),
\end{equation}
where $\beta$ controls the smooth maximum. A larger $\beta$ makes the log-sum-exp closer to the maximum but can concentrate gradients on a few grid samples. The sidelobe penalty is
\begin{equation}
	\mL_{\rm psll}
	=\operatorname{softplus}\left(
	\frac{20\log_{10}(\widehat P_{\rm sl}+\epsilon)-\Gamma_{\rm psll}}{2}
	\right),
\end{equation}
where $\Gamma_{\rm psll}$ is the target sidelobe level in dB. The training loss is
\begin{equation}
	\mL(\bm\theta)=-\E_{\rho_{\rm tr}}\left[R(\mathbf a)\right]
	+\lambda_{\rm psll}\mL_{\rm psll}
	+\lambda_c\mL_{\rm card}+\lambda_b\mL_{\rm bin},
	\label{eq:loss}
\end{equation}
where $\mL_{\rm card}$ penalizes deviation of the soft-mask mass from $M$, and $\mL_{\rm bin}$ discourages ambiguous soft activations. The surrogate is used only for training; all numerical PSLL results are computed from the final binary mask using \eqref{eq:psll_guard}. Thus, any continuous-mask approximation error affects learning gradients but not the reported PSLL values. During training, the SNR can be randomly perturbed as $\rho_{\rm tr,dB}\sim \mathcal U(\rho_{0,{\rm dB}}-\Delta_{\rm SNR},\rho_{0,{\rm dB}}+\Delta_{\rm SNR})$ to avoid fitting the scorer to a single operating SNR.

\subsection{Multi-Scale Geometry-Aware Inference}\label{subsec:geometry_inference}

At inference, L-BPA does not use raw block-wise top-$m_b$. Within each block, ports are selected sequentially using learned logits and a multi-scale spatial repulsion. Suppose $\mS_b^{(t)}$ is the set already selected in block $b$. For a candidate port $n\in\mB_b$, define
\begin{align}
	D_t(n)=\sum_{i\in\mS_b^{(t)}}\bigg[
	&\exp\left(-\frac{\|\mathbf p_n-\mathbf p_i\|^2}{2\sigma_1^2}\right) \\
	&+\rho_g\exp\left(-\frac{\|\mathbf p_n-\mathbf p_i\|^2}{2\sigma_2^2}\right)\bigg].
	\label{eq:density}
\end{align}
The next selected port is
\begin{equation}
	n^\star=\arg\max_{n\in\mB_b\setminus\mS_b^{(t)}}\left(\bar z_n-\mu_gD_t(n)\right),
	\label{eq:selector}
\end{equation}
where $\bar z_n$ is the normalized logit. The short scale $\sigma_1$ suppresses local clustering, while the larger scale $\sigma_2$ encourages spatial spreading inside each block. The process repeats until $m_b$ ports are selected from every block. The final mask is binary and exactly feasible.

The online computation consists of one neural forward pass, block-wise vector updates, and one RZF computation on the selected $K\times M$ channel. Unlike Greedy selection, L-BPA does not repeatedly solve RZF for many candidate port additions during inference.

\subsection{Online Complexity}\label{subsec:online_complexity}

The online L-BPA selector contains one neural forward pass and block-wise deterministic geometry selection. Let $C_\theta$ denote the forward-pass cost of the learned scorer. For block $b$, let $N_b$ and $m_b$ be the numbers of candidate and selected ports, respectively. The pairwise geometry kernel costs $O(N_b^2)$ and the sequential intra-block selection costs $O(m_bN_b)$, so the online selection complexity is
\begin{equation}
	O\!\left(C_\theta+\sum_{b=1}^{B}(N_b^2+m_bN_b)\right).
	\label{eq:blpa_complexity}
\end{equation}
After the active set is obtained, the common RZF backend on $\mathbf H_{\mS}\in\mathbb C^{K\times M}$ requires forming a $K\times K$ Gram matrix and solving a regularized linear system, with dominant cost $O(MK^2+K^3)$. This backend is shared by all communication-oriented methods in the simulations and is therefore separated from the L-BPA selection cost.
\begin{algorithm}[t!]
	\caption{Proposed L-BPA Training and Online Inference}
	\label{alg:L-BPA}
	\footnotesize
	\begin{algorithmic}[1]
		\REQUIRE Training channels $\{\mathbf H^{(i)}\}$, positions $\{\mathbf p_n\}$, blocks $\{\mB_b\}$, budgets $\{m_b\}$, trained scorer $f_{\bm\theta}$ for inference
		\ENSURE Feasible active set $\mS$ and RZF precoder $\widetilde{\mathbf P}_{\mS}$
		\STATE \textit{Offline training}
		\FOR{each mini-batch of channel realizations}
		\STATE Build $\mathbf X$ by \eqref{eq:feature}, draw $\rho_{\rm tr}$, and compute $\mathbf z=f_{\bm\theta}(\mathbf X)$.
		\FOR{each block $b$}
		\STATE Form $\mathbf a_b^{\rm soft}$ and $\mathbf a_b^{\rm hard}=\operatorname{TopM}(\mathbf z_b,m_b)$.
		\STATE Set $\mathbf a_b=\mathbf a_b^{\rm hard}+\mathbf a_b^{\rm soft}-\operatorname{sg}(\mathbf a_b^{\rm soft})$.
		\ENDFOR
		\STATE Concatenate $\mathbf a$, compute $R(\mathbf a)$ and $\mL_{\rm psll}$, and update $\bm\theta$ by \eqref{eq:loss}.
		\ENDFOR
		\STATE \textit{Online inference for a new CSI realization}
		\STATE Build $\mathbf X$, compute $\mathbf z=f_{\bm\theta}(\mathbf X)$, and normalize it to $\bar{\mathbf z}$.
		\FOR{each block $b$}
		\STATE Initialize $\mS_b=\emptyset$.
		\FOR{$t=1$ to $m_b$}
		\STATE Compute $D_t(n)$ for all $n\in\mB_b\setminus\mS_b$ by \eqref{eq:density}.
		\STATE Select $n^\star=\arg\max_{n\in\mB_b\setminus\mS_b}(\bar z_n-\mu_gD_t(n))$.
		\STATE Update $\mS_b\leftarrow\mS_b\cup\{n^\star\}$.
		\ENDFOR
		\ENDFOR
		\STATE Set $\mS=\cup_b\mS_b$ and compute $\widetilde{\mathbf P}_{\mS}$ from $\mathbf H_{\mS}$ using \eqref{eq:rzf}.
		\STATE Return $\mS$ and $\widetilde{\mathbf P}_{\mS}$.
	\end{algorithmic}
\end{algorithm}
\section{Simulation Setup and Numerical Results}\label{sec.4}

\subsection{Simulation Setup}

The FAA has $N_x=50$ and $N_y=25$ candidate ports over a $5\lambda\times2.5\lambda$ aperture, giving $N=1250$. The main sparse configuration activates $M=300$ ports, i.e., 24\% of the candidate ports. The downlink serves $K=16$ users. The channel generator follows a UMi setting at 3.5 GHz with BS height 10 m and UE height 1.5 m. For each user, the large-scale link distance is independently drawn from 50--150 m. The model uses 5 clusters and 12 rays per cluster. Therefore $L_k=60$ for NLoS users and $L_k=61$ when LoS is present. Cluster powers are exponentially distributed and normalized, NLoS ray phases are uniform over $[0,2\pi)$, LoS users follow a Rician component with $K$-factor drawn from a normal distribution in dB with mean 7 dB and standard deviation 4 dB, and shadowing/blockage are drawn before the per-user RMS normalization described in Section~\ref{sec.2}. This distance range is far beyond the Rayleigh distance of the $5\lambda$ aperture, so the far-field steering response in \eqref{eq:channel} is used.
\begin{table}[t!]
	\centering
	\caption{Main 500-sample performance at 10 dB. The 5\%-ile rate is the empirical lower-tail sum-rate over test channels. Bold values mark the full-port rate reference, the best sparse-method rate, and the lowest PSLL.}
	\label{tab:main_summary}
	\begin{tabular}{lccc}
		\toprule
		Method & Avg. rate & 5\%-ile rate & Avg. PSLL \\
		& (bit/s/Hz) & (bit/s/Hz) & (dB) \\
		\midrule
		\textbf{Full-port} & \textbf{129.06} & \textbf{115.26} & -13.21 \\
		\textbf{L-BPA} & \textbf{98.06} & \textbf{84.90} & -15.64 \\
		\textbf{Uniform} & 97.48 & 83.86 & -12.38 \\
		\textbf{Random} & 95.54 & 82.09 & -12.31 \\
		\textbf{Greedy} & 87.60 & 72.70 & -7.51 \\
		\textbf{PSLL-CGA} & 94.81 & 81.64 & -20.80 \\
		\textbf{PSLL-IGA} & 95.02 & 82.18 & \textbf{-21.42} \\
		\textbf{Gain-based} & 77.77 & 61.29 & -5.54 \\
		\bottomrule
	\end{tabular}
\end{table}
\begin{figure*}[t!]
	\centering
	\includegraphics[width=\linewidth]{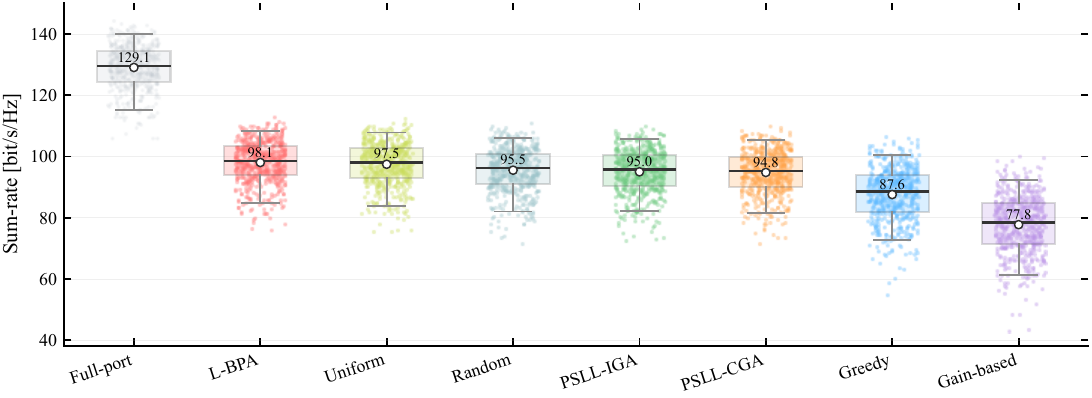}
	\caption{RZF sum-rate box plots over 500 test channels at 10 dB. Boxes show the interquartile range, center lines show medians, whiskers show the non-outlier range, and overlaid points show individual channel realizations. All sparse methods use $M=300$ active ports.}
	\label{fig:rate_boxplot}
\end{figure*}
\begin{figure*}[t!]
	\centering
	\includegraphics[width=\linewidth]{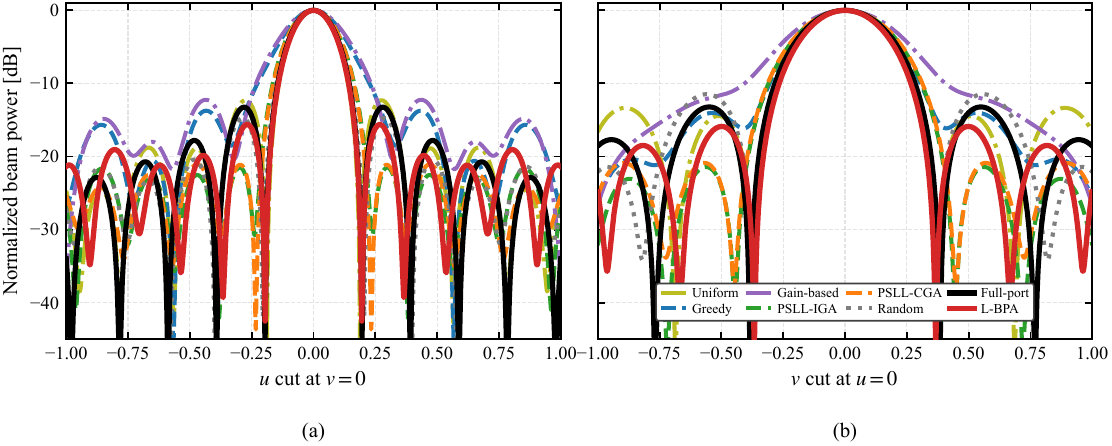}
	\caption{Normalized broadside beam-power cuts for a representative test sample: (a) cut along the $u$ axis with $v=0$; (b) cut along the $v$ axis with $u=0$. Curves are computed from equal-amplitude array factors and normalized to their visible-region maxima.}
	\label{fig:cuts}
\end{figure*}
\begin{figure*}[t!]
	\centering
	\includegraphics[width=\linewidth]{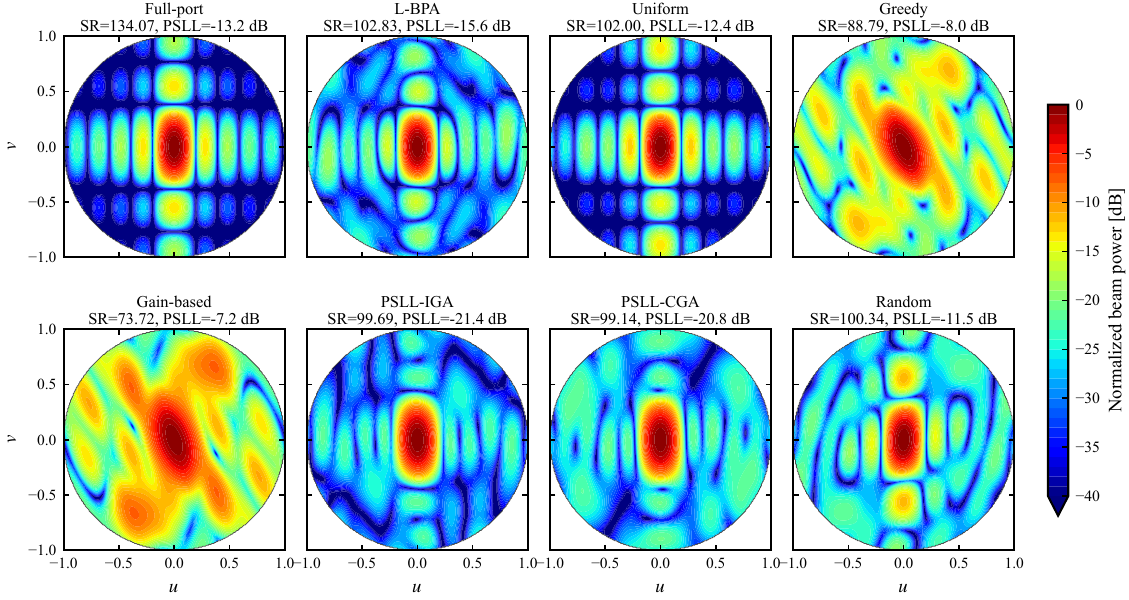}
	\caption{Two-dimensional normalized broadside beam-power maps for the same sample as Fig.~\ref{fig:cuts}. Each subfigure corresponds to one activation method; the circular support is the visible direction-cosine region $u^2+v^2\le1$, and the color scale is clipped to $[-40,0]$ dB.}
	\label{fig:uvmaps}
\end{figure*}
The direction-cosine quantization uses $N_u=50$ and $N_v=25$ grid points. The dataset contains 1200 training, 200 validation, and 500 test channel realizations. The neural scorer is trained for 60 epochs with AdamW \cite{Loshchilov2019AdamW}, learning rate $10^{-4}$, weight decay $10^{-5}$, batch size 8, hidden width 48, and Gumbel temperature $0.6$. The RZF regularization is fixed at $\eta_{\rm rzf}=0.1$ for all compared methods, so the study isolates port activation rather than tuning the precoder backend. In the geometry-aware selector, $\sigma_1$ and $\sigma_2$ are the short- and medium-range repulsion widths, $\rho_g$ controls the second-scale weight, and $\mu_g$ controls the global edge-dispersion term; their values are $\sigma_1=0.20\lambda$, $\sigma_2=0.35\lambda$, $\rho_g=0.5$, and $\mu_g=0.15$. The PSLL surrogate uses weight $\lambda_{\rm psll}=1.2$, target $\Gamma_{\rm psll}=-15$ dB, log-sum-exp sharpness $\beta=80$, and a $51\times51$ training grid; reported PSLL values are recomputed on hard binary masks using a denser $121\times121$ grid.

\begin{figure*}[t!]
	\centering
	\includegraphics[width=\linewidth]{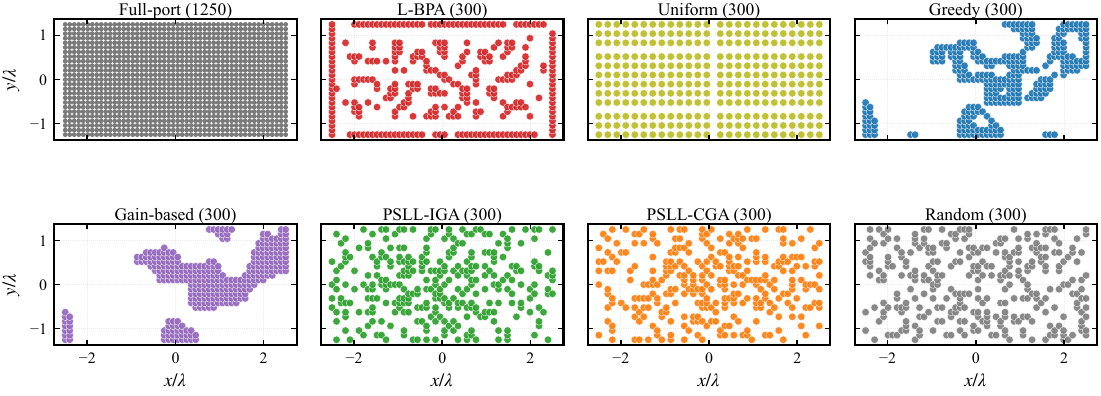}
	\caption{Active-port layouts for the representative sample used in Figs.~\ref{fig:cuts} and \ref{fig:uvmaps}. Each dot is an activated port on the $5\lambda\times2.5\lambda$ aperture. The number in parentheses is the active-port count.}
	\label{fig:ports}
\end{figure*}
\begin{figure*}[t!]
	\centering
	\includegraphics[width=\linewidth]{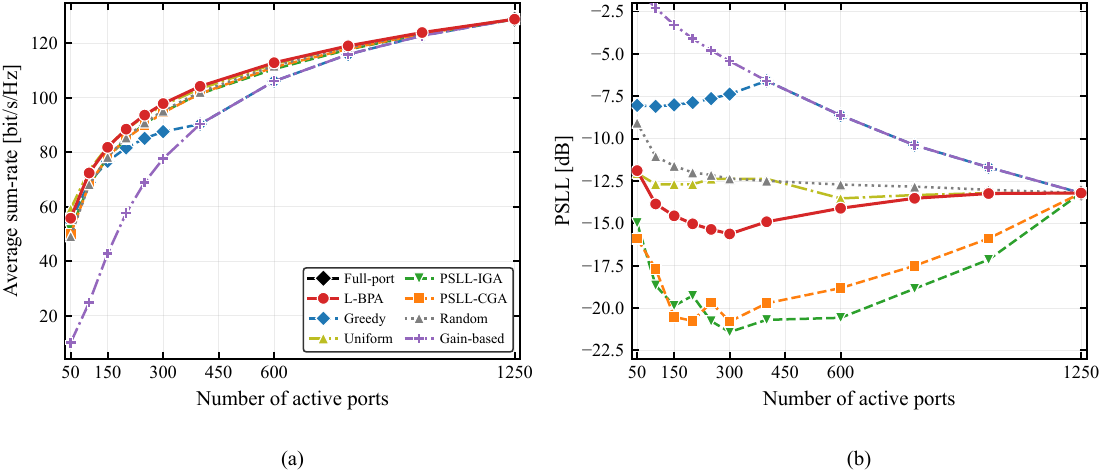}
	\caption{Active-port tradeoff between average RZF sum-rate and equal-amplitude PSLL over 200 test samples. Each marker corresponds to one active-port budget; the rightmost common endpoint is the full-port case with all $1250$ ports activated.}
	\label{fig:active_tradeoff}
\end{figure*}
The comparison includes the following methods.
\begin{itemize}
	\item \textbf{\textit{Full-port}}: all $1250$ ports are activated. It is an upper reference in port count, not a sparse method.
	\item \textbf{\textit{L-BPA}}: the proposed block-wise learned and geometry-aware activation method.
	\item \textbf{\textit{Uniform}}: a fixed rectangular sub-grid uniform sparse activation pattern.
	\item \textbf{\textit{Random}}: $M$ ports are drawn uniformly at random.
	\item \textbf{\textit{Gain-based}}: the $M$ ports with the largest aggregate multi-user channel power $\sum_k|h_{k,n}|^2$ are selected. This is a norm-based low-complexity antenna/port selection baseline.
	\item \textbf{\textit{Greedy}}: a communication-only greedy selector. Candidate ports are prefiltered by aggregate channel power, and ports are sequentially added to maximize the RZF sum-rate.
	\item \textbf{\textit{PSLL-CGA}} and \textbf{\textit{PSLL-IGA}}: genetic-algorithm sparse-array baselines \cite{beam2} that optimize broadside PSLL and do not use instantaneous multi-user channel utility.
\end{itemize}
All sparse methods use the same active-port budget when compared at $M=300$. All communication metrics are computed using the same RZF backend.
\subsection{Rate Distribution}

The 500-sample test results at 10 dB are summarized in Table~\ref{tab:main_summary}. Full-port gives the highest rate because it uses all ports. Among the sparse methods, L-BPA obtains 98.06 bit/s/Hz average sum-rate, higher than Uniform, Random, Greedy, PSLL-CGA, PSLL-IGA, and Gain-based in this evaluation. Its 5th-percentile rate is also the largest among the sparse methods. The PSLL values reveal the intended compromise: PSLL-only baselines have the lowest PSLL but sacrifice multi-user rate, whereas Greedy and Gain-based retain channel adaptivity at the cost of poor sidelobes. L-BPA improves the Uniform PSLL by 3.26 dB while preserving a slightly higher average rate.

Fig.~\ref{fig:rate_boxplot} shows the complete sum-rate distribution behind Table~\ref{tab:main_summary}. The L-BPA median and lower whisker are above Uniform and Random, so the rate gain is not caused by a small number of favorable channel realizations. The Greedy curve has lower lower-tail rate than L-BPA because the greedy search uses a restricted high-gain candidate pool and does not impose aperture coverage. Gain-based selection has the weakest distribution: it repeatedly selects ports with similar spatial responses, which reduces effective channel diversity after RZF even though the selected ports have large channel norms.
\begin{figure*}[t!]
	\centering
	\includegraphics[width=\linewidth]{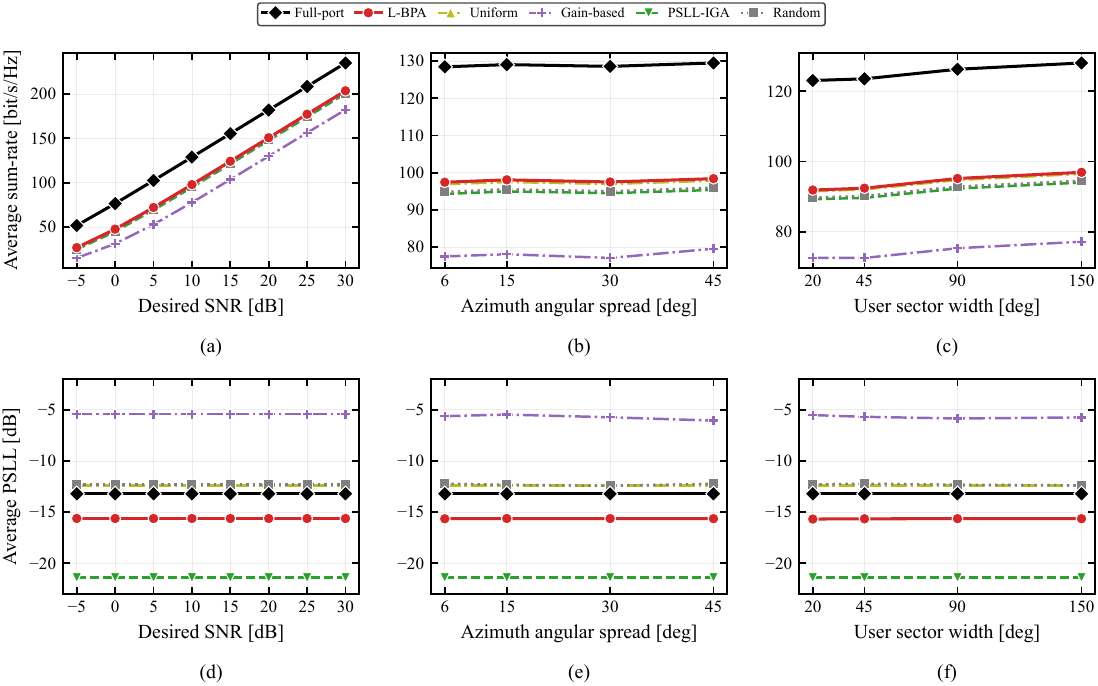}
	\caption{No-retraining generalization with $M=300$ active ports. (a) Average RZF sum-rate versus desired SNR. (b) Average RZF sum-rate versus azimuth angular spread. (c) Average RZF sum-rate versus user angular-sector width. (d) Average PSLL versus desired SNR. (e) Average PSLL versus azimuth angular spread. (f) Average PSLL versus user angular-sector width, where a smaller sector means more correlated user directions.}
	\label{fig:generalization}
\end{figure*}
\begin{table}[t]
	\centering
	\caption{Component ablation on the same $K=16$, $M=300$ distance-randomized test set as Table~\ref{tab:main_summary}.}
	\label{tab:ablation}
	\footnotesize
	\setlength{\tabcolsep}{0pt}
	\begin{tabular*}{\columnwidth}{@{\extracolsep{\fill}}lcccc@{}}
		\toprule
		Variant & Rate & 5\%-ile & PSLL & 95\%-ile \\
		& b/s/Hz & b/s/Hz & dB & dB \\
		\midrule
		\textbf{Random} & 95.54 & 82.09 & -12.31 & -10.66 \\
		\textbf{Uniform} & 97.48 & 83.86 & -12.38 & -12.38 \\
		\textbf{w/o learning} & 94.18 & 80.46 & -11.42 & -9.14 \\
		\textbf{w/o geom. loss} & \textbf{98.29} & 84.68 & -12.15 & -10.86 \\
		\textbf{Full L-BPA} & 98.06 & \textbf{84.90} & -15.64 & \textbf{-14.93} \\
		\bottomrule
	\end{tabular*}
\end{table}
\subsection{Beam Pattern and Port Distribution}

Fig.~\ref{fig:cuts} compares one-dimensional cuts through the broadside pattern along $v=0$ and $u=0$. Full-port provides the smooth reference aperture, while L-BPA follows it more closely than Greedy and Gain-based outside the mainlobe guard region. The large peaks of Greedy and Gain-based in both cuts are the visible consequence of clustered port selections. PSLL-CGA and PSLL-IGA give the lowest cuts because they optimize the equal-amplitude pattern directly; however, their masks are not conditioned on the instantaneous multi-user channel, which explains the rate loss in Table~\ref{tab:main_summary}.

For the same selected channel realization, Fig.~\ref{fig:uvmaps} presents the two-dimensional pattern over the visible direction-cosine disk $u^2+v^2\le1$. The white area is outside the physical visible region, not a zero-power region. Full-port has a regular broadside mainlobe and structured sidelobes. Gain-based and Greedy produce fragmented high-power sidelobe regions away from broadside because their active ports are spatially concentrated. L-BPA reduces these fragmented sidelobes while remaining less pattern-regular than the purely PSLL-driven masks, which is consistent with its role as a communication-aware compromise.

The port layouts in Fig.~\ref{fig:ports} explain the pattern behavior. Uniform forms a regular sparse grid, so it avoids severe clustering but cannot react to the channel. Random is dispersed on average but has no deterministic coverage guarantee for a given channel. Gain-based and Greedy concentrate many selected ports in high-gain regions, which shortens the effective aperture and causes the sidelobes seen in Fig.~\ref{fig:uvmaps}. The PSLL-oriented masks show a stronger tendency to occupy aperture edges and corners, because enlarging the effective aperture helps suppress broadside sidelobes for an equal-amplitude sparse array. L-BPA exhibits a milder version of this boundary-aware behavior: the block-wise constraint maintains aperture coverage, while the learned CSI scores prevent the selection from becoming a purely edge-dominated pattern.

\subsection{Active-Port Tradeoff}
Fig.~\ref{fig:active_tradeoff} traces the same active-port sweep from two complementary views. Fig.~\ref{fig:active_tradeoff}(a) reports the average RZF sum-rate versus the number of activated ports, while Fig.~\ref{fig:active_tradeoff}(b) reports the corresponding equal-amplitude PSLL; a desirable method should move upward in Fig.~\ref{fig:active_tradeoff}(a) without moving upward in Fig.~\ref{fig:active_tradeoff}(b). In the low- and medium-port regimes, Greedy and Gain-based selection recover rate by choosing strong channel samples, but their PSLL is high because the selected ports are spatially clustered. Uniform and Random give better aperture dispersion but do not exploit instantaneous CSI. L-BPA stays close to the high-rate sparse methods in Fig.~\ref{fig:active_tradeoff}(a) while keeping a substantially lower PSLL than the communication-only selectors in Fig.~\ref{fig:active_tradeoff}(b). At $M=1250$, all masks become the full-port aperture, so all curves meet at the same endpoint.

\subsection{Generalization Without Retraining}

No-retraining generalization is evaluated in Fig.~\ref{fig:generalization}. The figure uses a $2\times3$ layout: the top row reports average RZF sum-rate, the bottom row reports average PSLL, and the three columns sweep desired SNR, azimuth angular spread, and user angular-sector width, respectively. In Fig.~\ref{fig:generalization}(a), the desired SNR is changed only at evaluation. L-BPA follows the monotone rate increase of Full-port and Uniform, which indicates that the SNR perturbation used in training does not overfit the selector to a single operating SNR. In Fig.~\ref{fig:generalization}(b), increasing the azimuth angular spread creates richer multipath directions; the L-BPA rate changes mildly, showing weak sensitivity to this scattering mismatch. In Fig.~\ref{fig:generalization}(c), reducing the user angular-sector width makes user channels more correlated, so the sum-rate decreases for all communication methods. L-BPA cannot remove this fundamental multi-user correlation penalty, but it remains competitive among the sparse methods.

Figs.~\ref{fig:generalization}(d)--(f) report the corresponding PSLL behavior. In Fig.~\ref{fig:generalization}(d), PSLL is nearly flat across SNR because it is determined mainly by the selected geometry rather than the RZF power scale. In Fig.~\ref{fig:generalization}(e), L-BPA keeps a lower PSLL than Uniform and Gain-based over the angular-spread sweep, indicating that the learned geometric regularization is not tied to one nominal spread value. In Fig.~\ref{fig:generalization}(f), narrower user sectors change the communication channel correlation but do not strongly affect the equal-amplitude sparse-aperture PSLL. PSLL-IGA remains the strongest sidelobe-only method in these panels, but it gives up channel adaptivity and therefore does not dominate the rate plots in Figs.~\ref{fig:generalization}(a)--(c).

\subsection{Ablation Study}

The component ablation results in Table~\ref{tab:ablation} use the same distance-randomized $K=16$, $M=300$ test set as Table~\ref{tab:main_summary}. Replacing L-BPA with fixed Uniform or Random activation removes CSI-dependent learning; Uniform remains a strong geometry-only baseline, but its PSLL is worse than L-BPA. The \textbf{\textit{w/o learning}} variant removes the neural scorer and uses the explicit multi-user channel-power score $\sum_k|h_{k,n}|^2$ with the same block-wise geometry selector. Its rate and PSLL both degrade, showing that the learned score is not merely reproducing a norm-based rule. The \textbf{\textit{w/o geom. loss}} variant removes the training-stage geometric PSLL surrogate while keeping the learned scorer, block-wise mask, and inference rule unchanged. It gives the largest average rate, but the average PSLL degrades from $-15.64$ dB to $-12.15$ dB and the 95\%-ile PSLL becomes close to Random. Full L-BPA is therefore emphasized as the intended operating point: it preserves a high lower-tail rate while substantially improving sidelobe robustness.

\subsection{Discussion}

These results should not be interpreted as a sparse learned method outperforming full-port activation in sum-rate. Full-port remains the upper reference because it exposes the largest channel dimension to RZF. The relevant claim is different: with 24\% of the ports activated, L-BPA reaches 98.06 bit/s/Hz average sum-rate, corresponding to about 76.0\% of the full-port reference, while giving the strongest sparse lower-tail rate in Table~\ref{tab:main_summary}. Compared with Uniform, L-BPA improves average PSLL by 3.26 dB and increases the average rate slightly. Compared with Greedy and Gain-based selection, L-BPA reduces average PSLL by 8.13 dB and 10.10 dB, respectively, while also increasing the average rate by 10.46 and 20.28 bit/s/Hz. PSLL-IGA and PSLL-CGA give excellent sidelobe suppression, but they are static pattern-synthesis baselines after offline genetic optimization and do not adapt their mask to the instantaneous multi-user channel. L-BPA therefore occupies the intended operating point: it is channel-adaptive like Greedy, maintains substantially better sidelobes, and avoids online combinatorial search.

\section{Conclusion}\label{sec.5}
This paper proposed L-BPA scheme for real-time sidelobe-aware multiuser FAA downlink beamforming. L-BPA integrates channel-aware port scoring, deterministic block-wise constrained activation, and multi-scale geometry-aware selection to jointly balance sum rate and sidelobe suppression under a strict active-port budget. Straight-through block masks and a differentiable PSLL surrogate enable end-to-end training while preserving the discrete grouped switching constraints required at inference. The selected ports are combined with RZF precoding without iterative online search. Numerical results showed that L-BPA improves the average PSLL by 3.26 dB over uniform sparse activation while achieving a slightly higher sum rate. It further provides PSLL improvements of 8.13 dB and 10.10 dB over greedy and gain-based selection, respectively. 

These results demonstrate that channel awareness and geometric diversity can jointly improve radiation-pattern quality, communication performance, hardware compatibility, and online efficiency. Future work will consider wideband beam-squint effects and joint activation-precoding learning under imperfect CSI.

\balance

\end{document}